%% file: psip_rc.tex
\documentclass[prl,aps,superscriptaddress,altaffilletter]{revtex4}

\usepackage{graphicx}
\usepackage{epsfig}
\usepackage{amsfonts,amsmath}
\usepackage{dcolumn}
\usepackage{array} 

\newcommand{\dsdpt}{$\frac{d\sigma}{dp_T}\cdot$ Br}

\def\Journal#1#2#3#4{{#1} {\bf #2}, #3 (#4)}
\def \PRL      {Phys. Rev. Lett.}

\def \PRD      {Phys. Rev. D}

\def \NIMA      {Nucl. Instrum. Methods A}

\def \EPJC     {Eur. Phys. J. C}

\def \IEEE     {IEEE Trans. Nucl. Sci.}

\begin{document}

\title{ Production of $\psi$(2S) Mesons in $p\overline{p}$ Collisions at
  1.96 TeV}
\vspace*{.5cm}

\input{July_2008_Authors_Visitors}
\vspace*{.5cm}

\begin{abstract}
We have measured the differential cross section for the inclusive production of
$\psi$(2S) mesons decaying to $\mu^+\mu^-$ that were produced in 
prompt or $B$-decay processes from $p\overline{p}$ collisions at 1.96 TeV.  
These measurements have been made using a data set from an integrated 
luminosity of  1.1 fb$^{-1}$ collected by the CDF II detector at Fermilab.
For events with transverse momentum $p_T(\psi(\mathrm{2S})) >$ 2 GeV/$c$ 
and rapidity $| y(\psi(\mathrm{2S})) | <$ 0.6 we measure 
the integrated inclusive cross section $\sigma(p\overline{p} \rightarrow \psi(\mathrm{2S}) X) \cdot \rm{Br} (\psi(\mathrm{2S})\rightarrow\mu^+ \mu^-)$ to be
$ 3.29 \pm 0.04 (stat.) \pm 0.32 (syst.)$ nb.
\end{abstract}

\maketitle

The mechanism for producing heavy vector mesons in $p\overline{p}$ 
collisions is not well understood.
The experimental measurement of prompt J/$\psi$ and $\psi$(2S) production 
cross sections by CDF in Tevatron Run I~\cite{run1-xs} showed that 
the measured cross sections were one to two orders of magnitude larger than 
expected from the leading order (LO) color-singlet models.
Theoretical efforts to improve the calculations added color octet 
contributions that increased the predicted cross sections, {\it e.g.}, 
in the non-relativistic QCD model (NRQCD)~\cite{nrqcd}.  
Recently there have been other approaches that do
not directly introduce a color octet amplitude; rather, they incorporate the 
effects of multiple gluon processes during the production process, {\it e.g.}, 
the $k_{T}$-factorization formalism~\cite{kt-factorization} 
and the gluon tower model~\cite{gluon-tower}.

The NRQCD model with parametrized production matrix elements adjusted to data 
can successfully account for the Tevatron prompt $\psi$(2S) cross section 
measurements, 
but it makes an unequivocal prediction of increasing transverse polarization 
of vector mesons as their transverse momentum $p_T$ from production 
increases~\cite{nrqcd}.  
A recent polarization measurement at CDF~\cite{run2-pol} contradicts 
the NRQCD model prediction. 

Experimentally, the extraction of direct J/$\psi$ production information is 
complicated by significant feed-down from decays of promptly-produced 
higher-mass charmonium states ($\chi_{c}, \psi$(2S)) to J/$\psi$ mesons.  
This is not a problem for direct $\psi$(2S) production because there are no 
reported charmonium states with significant hadronic production cross 
sections that decay to the $\psi(\mathrm{2S})$. 
Consequently the $\psi$(2S) provides an ideal testing ground for studying 
charmonium hadroproduction mechanisms.  
In this paper we present a measurement of the $p_T$ dependence of 
the $\psi$(2S) production cross section over 
the $\psi$(2S) transverse momentum range 
$2< p_T(\psi(\mathrm{2S})) < 30$ GeV/$c$ 
with rapidity $\mid y(\psi(\mathrm{2S})) \mid < 0.6$.
This measurement greatly increases the statistical power of the data in 
the perturbative regime ($m_T \gg \lambda_{QCD}$), 
facilitating comparison with theory.

We use data taken using the CDF II detector at the 
Fermilab Tevatron at 1.96 TeV~\cite{CDF}.  
The integrated luminosity of the data sample is 1.1 fb$^{-1}$. 
The CDF II detector, described in detail elsewhere~\cite{run2-xs}, 
includes a tracking system in a solenoidal 1.4 T magnetic field.
Electromagnetic and hadronic calorimeters backed by muon detectors 
surround the tracker. 
The essential detector elements for this analysis are 
the silicon strip tracking detector (SVX II), 
the central drift chamber (COT) and the central muon system (CMU and CMP).
The CMU is a four-layer planar drift chamber system outside 
the CDF magnet coil and calorimeter steel (5 interaction lengths).  
The CMP is another muon chamber system behind the CMU, 
shielded by an additional 0.6 m of iron in the flux return yoke.  
In this analysis, we use only information provided by the central 
sector of the detector, with pseudorapidity $\mid \eta \mid < 0.6$.

Muon candidates are identified by a first-level hardware-based trigger 
that reconstructs a charged track in four axial layers of the COT
~\cite{xft}.   The trigger then projects the track into the CMU/CMP 
system and matches the projected 
trajectory to a collection of three or four hits in the CMU muon system 
within a search window around the extrapolated track~\cite{xtrp}.  
The dimuon trigger requires two opposite-sign muon candidates each having  
$p_T > 1.5$ GeV/$c$. 

The $\psi(\mathrm{2S})\rightarrow \mu^+ \mu^-$ candidates were reconstructed 
from muon pairs.  The $\psi$(2S) events may originate from 
the primary interaction (prompt) or from decays of $B$-hadrons ($B$-decay).
In offline reconstruction, each muon had to have at least three hits in the 
$r\mbox{\small -} \phi$ strips of SVX II in order to guarantee good vertex 
information to separate prompt and $B$-decay candidates.  
The minimum muon $p_T$ is 2 GeV/$c$. 
If the CMU candidate has matching hits in the CMP chambers~\cite{run2-xs}, 
the track $p_T$ requirement is raised to 3 GeV/$c$ 
to account for the extra iron traversed.

The $\psi$(2S) mass and proper time distributions are used in a joint
unbinned maximum likelihood fit to extract the prompt and $B$-decay signals
in bins of $p_T$ for $\psi$(2S) candidates.  
The mass component separates signal from background, while the proper time 
component separates prompt $\psi$(2S) events from those produced by B-decays.
The mass distribution including the radiative tail is described by 
a combination of a Gaussian 
plus an asymmetric function (CBF)~\cite{CB} given by

\begin{equation*}
CBF = \left\{
\begin{array}{l l}
  A\cdot e^{-\frac{(M-m_{0})^{2}}{2\sigma^{2}}} 
& \ \mbox{$\frac{M-m_{0}}{\sigma}>-\beta$}\\
  A\cdot \left(\frac{n}{\beta}\right)^{n}
      \frac{e^{-\frac{\beta^{2}}{2}}}{\left(\frac{M-m_{0}}{\sigma}+
\frac{n}{\beta}-\beta \right)^{n} }
& \ \mbox{$\frac{M-m_{0}}{\sigma}\leq-\beta $}\\
  \end{array} \right.
\end{equation*}  
 where $m_0$ is a fit parameter for the invariant mass peak, 
$M$ is the dimuon invariant mass of each event, 
$A$ is the normalization constant, 
and empirical parameters $\beta$ and $n$ describe the tails of the function.

The parameters $n$ and $\beta$ of the tail function, and the relative fraction 
of the Gaussian and CBF are fixed by a fit to the 
entire $p_T$ range. 
The Gaussian and the Gaussian part of CBF have 
the same width $\sigma$.
This width is 
$p_T$-dependent due to experimental effects.  We use simulation results
to describe the relative $p_T$ dependence in the different bins, 
leaving one width parameter to be fit in the mass probability density function (PDF). 
The background mass function is linear ($P^{mass}_{bkg}\propto M$).  

The proper decay length $ct$ is used to identify prompt and $B$-decay 
contributions to the mass signal. Here $ct = \frac{L_{xy}}{p_T/M}$, 
where $L_{xy}$ is the transverse decay length projected onto 
the $\psi$(2S) momentum.  
The prompt component is described by a double Gaussian 
function centered at zero ($P^{ct}_{p}$).  
The long-lived component is an exponential ($P^{ct}_{long}$).
Because the $\psi$(2S) events 
from $B$-decay come from $B_h\rightarrow \psi(\mathrm{2S})X$, 
they do not have a $B$-hadron lifetime distribution. 
We use the effective lifetime of the $B$-decay signal as a fit parameter. 
Because it is defined in the $\psi$(2S) rest frame, 
it is the same in all $p_T$ bins.  
Finally, the background in the $ct$ distribution is described by the sum of 
a prompt double Gaussian ($P^{ct}_{pb}$) plus three exponentials, each
convolved with a Gaussian resolution function: 
one symmetric about zero ($P^{ct}_{sym}$), 
one for positive $ct$ only ($P^{ct}_{+}$), 
and one for negative $ct$ only ($P^{ct}_{-}$).  
The likelihood function is
\begin{eqnarray}
\lefteqn{ L = f_s P_s^{mass}[f_p P_p^{ct} + (1 - f_p) P_{long}^{ct}] + ( 1 - f_s)P_{bkg}^{mass}\times } \nonumber\\
& &  [f_{sym}P_{sym}^{ct} + f_+ P_{+}^{ct}+ f_- P_{-}^{ct} + (1 - f_{sym} - f_+ - f_-) P_{pb}^{ct}] \ . 
\label{eqn:likelihood}
\end{eqnarray}
The population fractions include
  $f_{s}$, the $\psi$(2S) signal fraction from the total number of 
	candidates in the fit,
  $f_{p}$, the fraction of prompt $\psi$(2S),
  $f_{sym}$, the fraction of symmetric long-lived background,
  $f_{+}$, the fraction of positive-ct long-lived background,
 and  $f_{-}$, the fraction of negative-ct long-lived background.
The PDFs for the mass fit include  
    $P^{mass}_{s}$, a  (CBF + Gaussian) distribution for signal,  
    and $P^{mass}_{bkg}$. 

\begin{figure}[htb]
 \centering
   \includegraphics[width=0.3\textwidth]{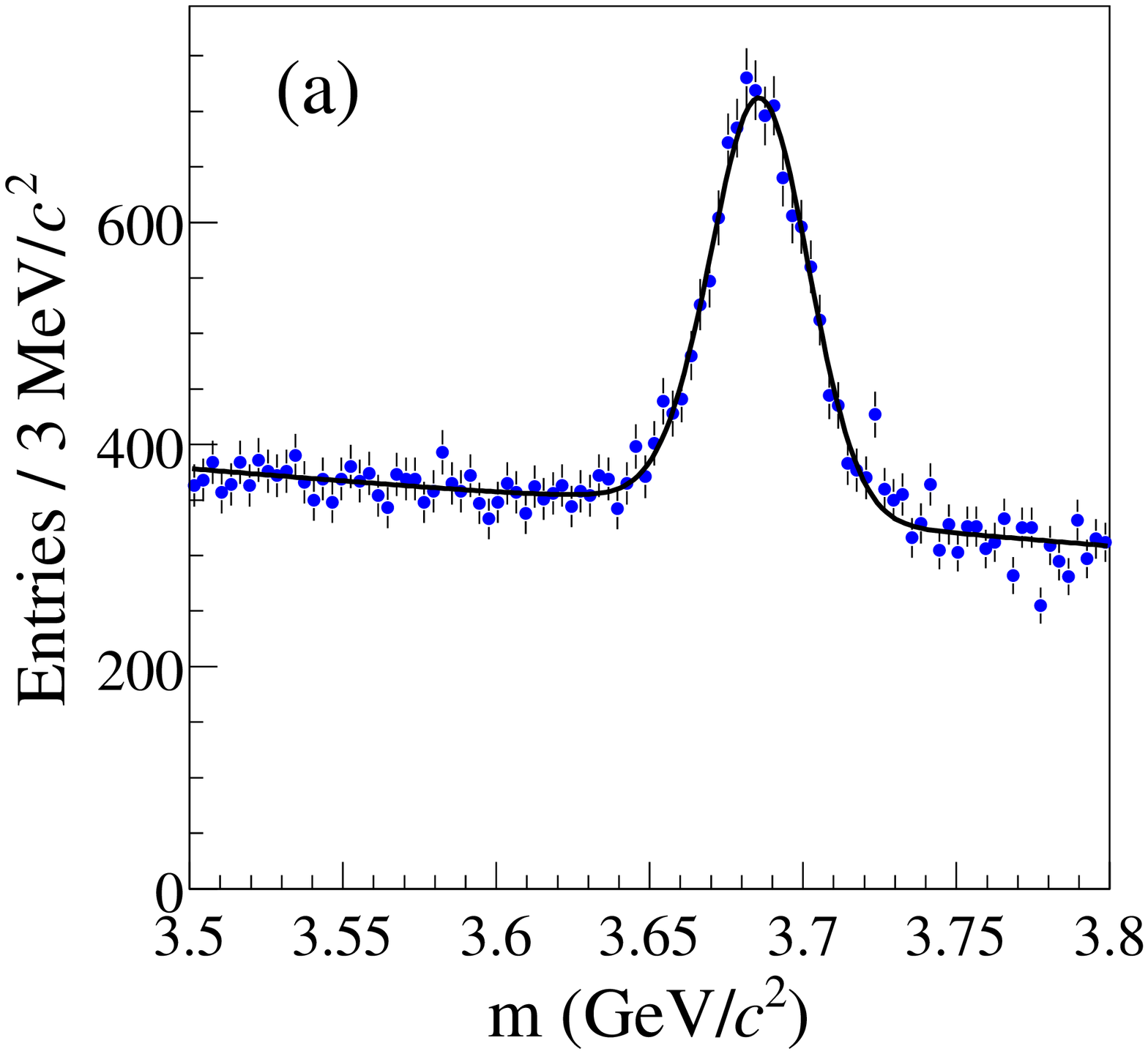}
   \includegraphics[width=0.3\textwidth]{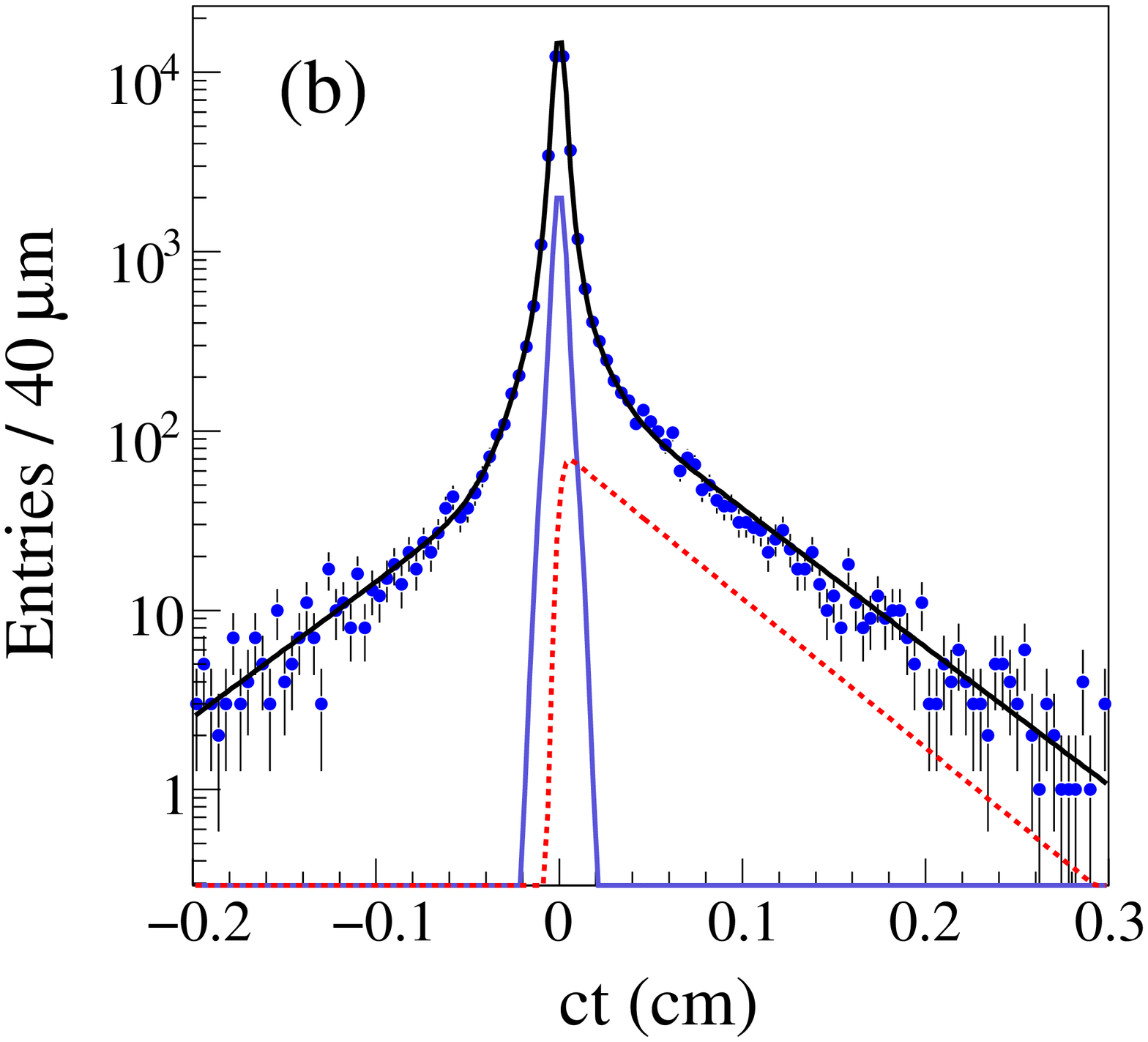}
 \caption{The projections on the invariant mass (a) and the proper decay 
          length (b) for $5.5 < p_T < 6.0$ GeV/$c$. 
          The fitted curve for total signal plus background is overlaid on
          the data points.  The prompt (solid) and long-lived (dashed) proper
          time signal curves are shown separately. \label{fig:fit-rep}}
\end{figure}

The fit parameters are the width of the overall Gaussian in $P_s^{mass}$,
the parameters of the double Gaussians for the prompt signal and background
$ct$ density distributions, the lifetime of the overall long-lived signal,
the five data fractions $f_i$ from Eq.(~\ref{eqn:likelihood}), 
and the parameters of the long-lived background functions in each $p_T$ bin. 
The fit projections in mass and proper time for $5.5 < p_T < 6.0$ GeV/$c$ 
are shown in  Fig.~\ref{fig:fit-rep}.   
Projecting the likelihood fits onto the mass distribution in each $p_T$ bin 
gives $\chi^2$ probabilities in the range $[0.4 - 100]\%$.   
The fitted yields in each $p_T$ bin are summarized in Table~\ref{tab1}.  
Signal events are classified as prompt or long-lived by 
the fit parameter $f_p$.  
For each $\psi{\mathrm(2S)}$ $p_T$  bin we know the total number of events $N$.
The fit returns the signal fraction $f_s$ and its uncertainty $\sigma_{f_s}$.
The signal yield is $S=f_s\cdot N$; $\sigma_{S}^2=(\sigma_{f_s}\cdot N)^2 +N\cdot f_s^2$.
Analogous equations hold for the prompt yield $P=f_p\cdot S$ and
the $B$-decay yield $B=(1-f_s)\cdot S$.
The correlation between the signal fraction and the prompt fraction
is considered in the uncertainty of the prompt and $B$-decay yield.
The number of prompt and $B$-decay events are also listed in Table~\ref{tab1}.

\begin{table}[htb]
\begin{scriptsize}
\centering
\begin{tabular}{c@{ $-$ }c r r@{ $\pm$ }r r@{ $\pm$ }r c r@{ $\pm$ }r}
\hline\hline
\multicolumn{2}{c}{$p_T$} & $\langle p_T \rangle$ &\multicolumn{2}{c}{Signal} &\multicolumn{2}{c}{Prompt} &   Prompt
fraction &\multicolumn{2}{c}{$B$-decay} \\
\hline
 2.0  & 2.5  &  2.3  & 1961 &  99 & 1701 &  96 & 0.867 $\pm$ 0.019 &  260 & 39  \\
 2.5  & 3.0  &  2.8  & 5025 & 157 & 4241 & 152 & 0.844 $\pm$ 0.011 &  785 & 58  \\
 3.0  & 3.5  &  3.2  & 7003 & 177 & 5955 & 170 & 0.850 $\pm$ 0.009 & 1048 & 64  \\
 3.5  & 4.0  &  3.8  & 6902 & 171 & 5754 & 166 & 0.834 $\pm$ 0.009 & 1148 & 63  \\
 4.0  & 4.5  &  4.2  & 7060 & 160 & 5778 & 153 & 0.818 $\pm$ 0.008 & 1282 & 60  \\
 4.5  & 5.0  &  4.7  & 6612 & 147 & 5376 & 141 & 0.813 $\pm$ 0.008 & 1236 & 57  \\
 5.0  & 5.5  &  5.2  & 5519 & 133 & 4462 & 127 & 0.809 $\pm$ 0.009 & 1057 & 52  \\
 5.5  & 6.0  &  5.7  & 5236 & 121 & 4213 & 114 & 0.805 $\pm$ 0.009 & 1023 & 50  \\
 6.0  & 6.5  &  6.2  & 4663 & 111 & 3636 & 108 & 0.780 $\pm$ 0.011 & 1027 & 51  \\
 6.5  & 7.0  &  6.7  & 3961 &  99 & 3105 &  94 & 0.784 $\pm$ 0.011 &  857 & 45  \\
 7.0  & 7.5  &  7.2  & 3173 &  87 & 2408 &  81 & 0.759 $\pm$ 0.012 &  765 & 41  \\
 7.5  & 8.0  &  7.7  & 2735 &  78 & 2066 &  73 & 0.756 $\pm$ 0.013 &  668 & 37  \\
 8.0  & 8.5  &  8.2  & 2209 &  69 & 1589 &  62 & 0.720 $\pm$ 0.014 &  619 & 35  \\
 8.5  & 9.0  &  8.7  & 1804 &  62 & 1261 &  56 & 0.699 $\pm$ 0.016 &  543 & 32  \\
 9.0  & 9.5  &  9.2  & 1418 &  55 &  987 &  49 & 0.696 $\pm$ 0.019 &  430 & 29  \\
 9.5  & 10   &  9.7  & 1170 &  50 &  800 &  45 & 0.684 $\pm$ 0.021 &  369 & 27  \\
 10   & 11   & 10.5  & 1692 &  60 & 1134 &  54 & 0.670 $\pm$ 0.018 &  558 & 33  \\
 11   & 12   & 11.5  & 1206 &  51 &  810 &  45 & 0.672 $\pm$ 0.021 &  395 & 28  \\
 12   & 13   & 12.5  &  788 &  41 &  511 &  36 & 0.648 $\pm$ 0.026 &  277 & 23  \\
 13   & 14   & 13.5  &  560 &  35 &  331 &  30 & 0.591 $\pm$ 0.032 &  229 & 21  \\
 14   & 15   & 14.5  &  410 &  29 &  240 &  25 & 0.586 $\pm$ 0.036 &  170 & 17  \\
 15   & 17.5 & 16.1  &  519 &  36 &  284 &  30 & 0.547 $\pm$ 0.036 &  235 & 22  \\
 17.5 & 20   & 18.6  &  242 &  26 &  129 &  22 & 0.535 $\pm$ 0.058 &  112 & 15  \\
 20   & 25   & 22.1  &  202 &  25 &  117 &  22 & 0.577 $\pm$ 0.063 &   86 & 14  \\
 25   & 30   & 27.1  &   74 &  17 &   45 &  14 & 0.609 $\pm$ 0.106 &   29 & 9   \\
\hline\hline
\end{tabular}
\end{scriptsize}
\caption{Event yields from the unbinned maximum likelihood fit and prompt event fraction. Transverse momenta are in GeV/$c$. The uncertainties are statistical only.}\label{tab1}
\end{table}

We have checked the $p_T$ dependence of all the fit parameters.  
The variation is smooth and shows no indications of rapid changes of 
the background functions at any $p_T$.  
The prompt fraction decreases approximately linearly
in the interval $2 < p_T < 30$ GeV/$c$.

The differential cross section is evaluated using the expression
\begin{equation}
\frac{d\sigma(\psi(\mathrm{2S}))}{dp_T} = \frac{N(\psi(\mathrm{2S}))}{\mathcal{A} \cdot 
\varepsilon_{reco}  \cdot \int \mathcal{L} dt \cdot \Delta p_T} \ \ .
\end{equation}

Here $\frac{d\sigma(\psi(\mathrm{2S}))}{dp_T}$ is the average cross section 
for  $\psi$(2S) production in the given $p_T$ bin integrated over rapidity in 
the range $\mid y \mid \le$ 0.6, 
$N(\psi(\mathrm{2S}))$ is the number of $\psi(\mathrm{2S})$ 
events determined by the fit, 
$\mathcal{A}$ is the geometric acceptance
combined with the CDF dimuon trigger efficiency, $\varepsilon_{reco}$ is the
reconstruction efficiency, $\int \mathcal{L} dt$ is the integrated luminosity
of the data set, and $\Delta p_T$ is the width of the $p_T$ bin.  
The acceptance and reconstruction efficiency are determined as follows.

The geometric acceptance is calculated by a Monte Carlo 
simulation (MC) method, 
using $\psi(\mathrm{2S})\rightarrow \mu^+ \mu^-$ decays generated uniformly 
for 1 $< p_T < 40$ GeV/$c$, 
$\mid y \mid < 1$, and $0 \le \phi \le 2\pi$.  The
$\psi(\mathrm{2S})$ decays are handled by {\sc evtgen}~\cite{evtgen}, allowing us 
to specify the decay polarization as transverse, longitudinal or
unpolarized.  We generate independent MC sets of these three options.
Because the tracking proceeds from the large-radius detectors inward, 
the geometric acceptance calculated for the prompt events is insensitive to 
small displacements of the $\psi$(2S) decay point. Therefore we use the 
same MC samples for calculating the geometric acceptance for both 
prompt and  $B$-decay events.  The systematic uncertainty for this 
assumption is negligible.

The MC events are passed through the CDF II {\sc geant}-based 
simulation~\cite{cdfsim} and the standard CDF reconstruction.
Events that pass the geometric selection are accepted based on 
each event's dimuon trigger efficiency, derived from CDF data for 
muon pairs having $\mid y \mid \le 0.6$ with 
each muon having $p_T \ge$ 2 GeV/$c$~\cite{bmumu}.
Variations with run and luminosity are included in the measurements.
The prompt MC sample was analyzed with the likelihood fitter 
to check for $p_T$ variations in prompt selection efficiency.  None were seen. 

Determining $\mathcal{A}$ 
is sensitive to the $\psi$(2S) polarization parameter $\alpha$, which 
defines the muon decay angular distribution in the vector 
meson rest frame: $ dN/d\cos\theta = 1+\alpha \cos^2\theta$. 
The polar angle $\theta$ is measured from the vector meson's direction 
in the laboratory frame.  

We have previously measured the $\psi$(2S) polarization in three $p_T$ bins 
for prompt events and the average polarization in $B$-decay 
events~\cite{run2-pol}.  By symmetry prompt events have $\alpha=0$ at $p_T=0$.  With 15\% probability a 
$\chi^2$ fit shows that the three measured points are consistent with 
$\alpha=0$. We use this as a basis to make the assumption that the 
polarization parameter $\alpha$ is constant over the $p_T$ range of 
the data.  Averaging the three measured points gives an average parameter 
$\alpha=0.01\pm 0.13$, which is used to determine $\mathcal{A}$
and its polarization-dependent 
systematic uncertainty. The prompt acceptance
$\mathcal{A}$ varies from 2\% at $p_T$ = 3 GeV/$c$ to 20\% at $p_T$ = 23 GeV/$c$.

For $B$-decay events we use the same procedure.  The polarization dependence is 
calculated using the measured $B$-decay polarization $\alpha_{eff} = 0.36 
\pm 0.25 \pm 0.03$~\cite{run2-pol}.  The $B$-decay acceptance varies from 1.5\% 
at $p_T$ = 3 GeV/$c$ to 19\% at $p_T$ = 23 GeV/$c$.  Since the polarization is 
different for the prompt and 
$B$-decay events, a weighted average of the acceptances in each $p_T$ bin is 
used for the inclusive differential cross section.

The reconstruction efficiency is the product of tracking and muon
selection efficiencies measured in CDF data, 
including the tracking efficiencies for the COT ($0.996\pm0.009$),
SVX II ($0.958 \pm 0.006$) and the dimuon 
tracking and selection efficiency ($0.875 \pm 0.019$).
Combining all the factors and adding the uncertainties in
quadrature gives $\varepsilon_{reco} = 0.805 \pm 0.038$.

\begin{table*}[t]
\begin{center}
\begin{footnotesize}
\begin{tabular}{c@{ $-$ }c @{\hspace{3ex}} c@{ $\pm$ }c@{ $\pm$ }c @{\hspace{6ex}} c@{ $\pm$ }c@{ $\pm$ }c @{\hspace{6ex}} c@{ $\pm$ }c@{ $\pm$ }c }
\hline\hline
\multicolumn{2}{c}{$p_T$} & \multicolumn{3}{c}{Inclusive \dsdpt} & \multicolumn{3}{c}{Prompt \dsdpt} & \multicolumn{3}{c}{$B$-decay \dsdpt} \\    
\multicolumn{2}{c}{(GeV/$c$)} & \multicolumn{3}{c}{(pb/GeV/$c$)} & \multicolumn{3}{c}{(pb/GeV/$c$)} & \multicolumn{3}{c}{(pb/GeV/$c$)} \\    
\hline 
2.0  & 2.5  & 1144     & 58     & 132      & 953     & 54     & 113     &  191     & 28     & 45     \\
2.5  & 3.0  & 1153     & 36     & 112      & 946     & 34     &  95     &  207     & 15     & 32     \\
3.0  & 3.5  & 1037     & 26     & 102      & 856     & 24     &  87     &  181     & 11     & 30     \\
3.5  & 4.0  &  779     & 19     &  73      & 630     & 18     &  61     &  149     &  8     & 23     \\
4.0  & 4.5  &  611     & 14     &  60      & 483     & 13     &  49     &  128     &  6     & 21     \\
4.5  & 5.0  &  490     & 11     &  48      & 383     & 10     &  39     &  107     &  5     & 18     \\
5.0  & 5.5  &  316     &  8     &  29      & 248     &  7     &  23     &   68     &  3     & 10     \\
5.5  & 6.0  &  262     &  6     &  24      & 204     &  6     &  19     &   58     &  3     &  8     \\
6.0  & 6.5  &  189     &  5     &  17      & 143     &  4     &  13     &   46     &  2     &  6     \\
6.5  & 7.0  &  146     &  4     &  13      & 111     &  3     &  10     &   35     &  2     &  5     \\
7.0  & 7.5  &  105.4   &  2.9   &   9.5    &  77.1   &  2.6   &   7.1   &   28.3   &  1.5   &  3.8   \\
7.5  & 8.0  &   81.9   &  2.3   &   7.4    &  59.8   &  2.1   &   5.5   &   22.1   &  1.2   &  2.9   \\
8.0  & 8.5  &   60.9   &  1.9   &   5.5    &  42.1   &  1.7   &   3.9   &   18.8   &  1.1   &  2.5   \\
8.5  & 9.0  &   45.4   &  1.6   &   4.0    &  30.5   &  1.3   &   2.7   &   14.9   &  0.9   &  1.8   \\
9.0  & 9.5  &   32.3   &  1.3   &   2.8    &  21.8   &  1.1   &   1.9   &   10.5   &  0.7   &  1.2   \\
9.5  & 10   &   25.9   &  1.1   &   2.2    &  17.2   &  1.0   &   1.5   &    8.7   &  0.6   &  0.9   \\
10   & 11   &   17.7   &  0.6   &   1.5    &  11.4   &  0.54  &   1.0   &    6.3   &  0.4   &  0.7   \\
11   & 12   &   11.96  &  0.5   &   1.1    &   7.72  &  0.43  &   0.69  &    4.24  &  0.30  &  0.51  \\
12   & 13   &    7.33  &  0.39  &   0.64   &   4.56  &  0.33  &   0.40  &    2.77  &  0.23  &  0.31  \\
13   & 14   &    5.03  &  0.32  &   0.44   &   2.85  &  0.26  &   0.24  &    2.18  &  0.20  &  0.24  \\
14   & 15   &    3.42  &  0.25  &   0.29   &   1.93  &  0.20  &   0.16  &    1.49  &  0.15  &  0.15  \\
15   & 17.5 &    1.61  &  0.11  &   0.14   &   0.85  &  0.09  &   0.07  &    0.76  &  0.07  &  0.08  \\
17.5 & 20   &    0.68  &  0.07  &   0.06   &   0.35  &  0.06  &   0.03  &    0.33  &  0.05  &  0.03  \\
20   & 25   &    0.27  &  0.03  &   0.02   &   0.15  &  0.03  &   0.01  &    0.12  &  0.02  &  0.01  \\
25   & 30   &    0.089 &  0.020 &   0.007  &   0.053 &  0.017 &   0.004 &    0.036 &  0.011 &  0.003 \\
\hline\hline
\end{tabular}
\end{footnotesize}
\end{center} 
\caption{The differential cross section (pb/GeV/$c$) times the dimuon branching 
fraction as a function of $p_T$ for $\mid y \mid \le0.6$.  For the $B$-decay measurement the symbol Br includes the branching fraction for b quark inclusive 
decay to $\psi$(2S)X as well as the dimuon branching fraction of $\psi$(2S). 
\label{xstab}}
\end{table*}

\begin{figure}[h]
\includegraphics[width=0.3\textwidth]{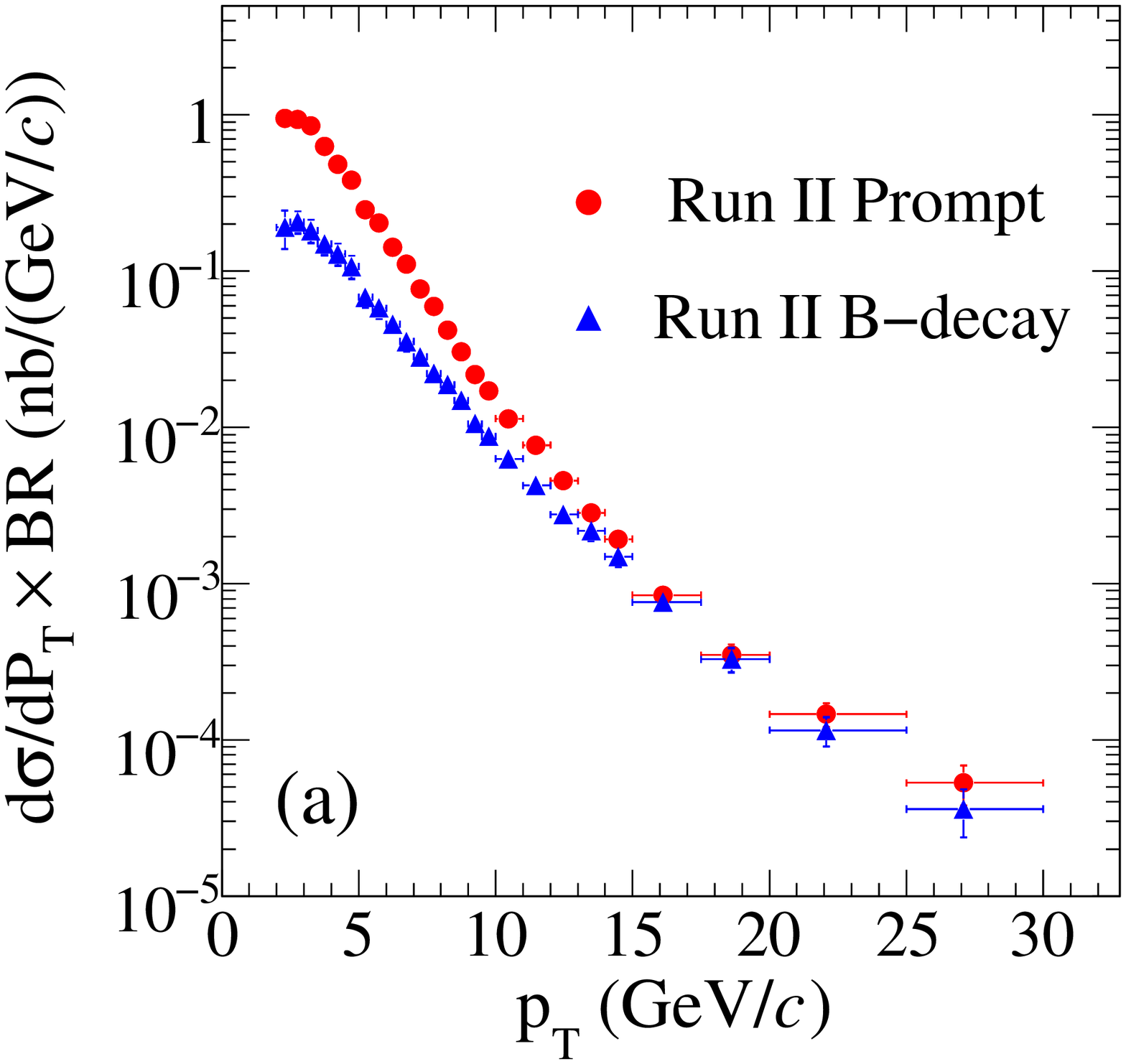}
\includegraphics[width=0.3\textwidth]{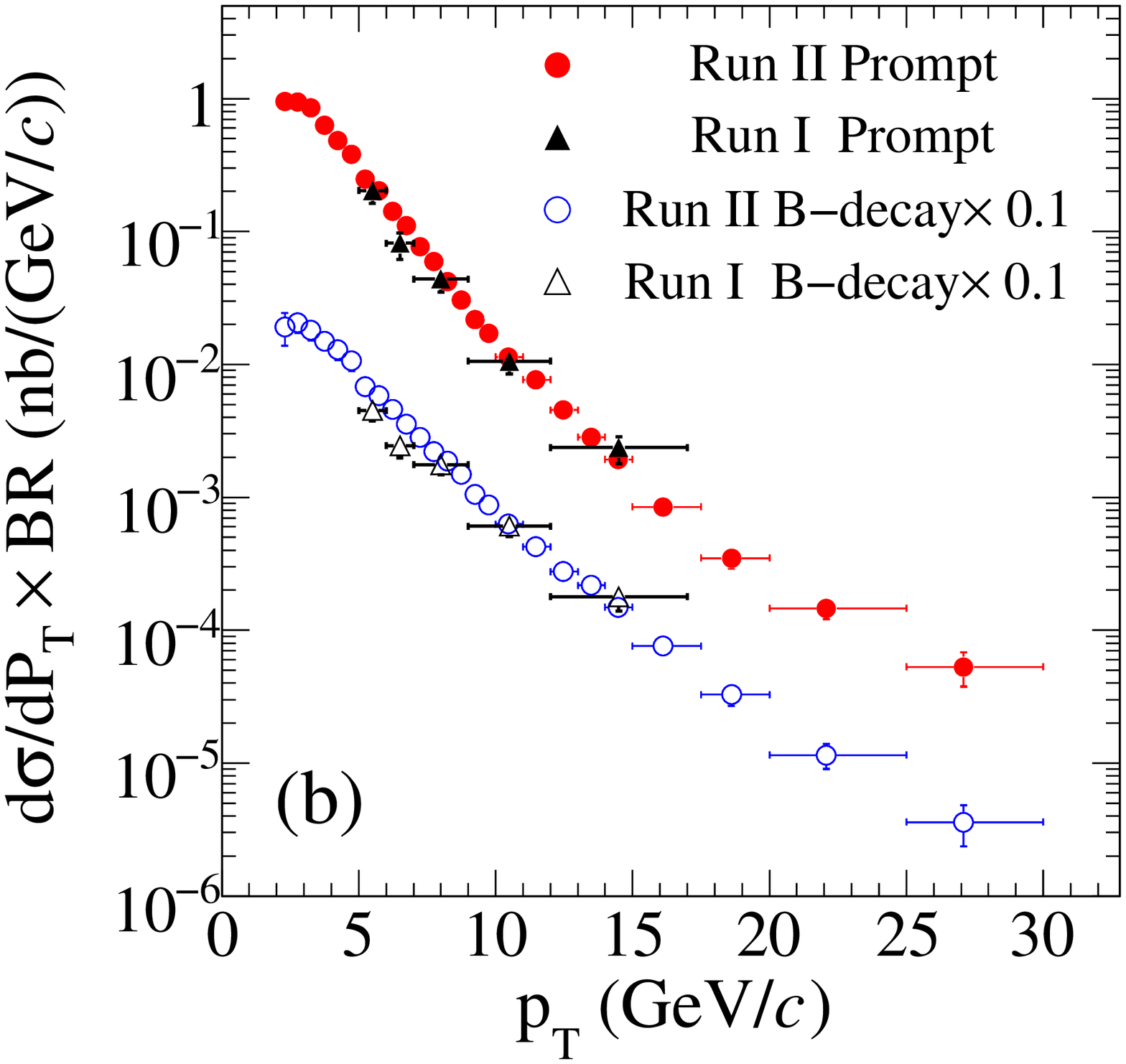}
\caption{(a): Prompt and $B$-decay production cross section distributions 
versus $p_T$ for these data. \ (b): The same data with the Run I points
included. We ignore differences between rapidity and pseudorapidity for this comparison.
The $B$-decay points have been scaled down by a factor of ten 
for clarity of display.}
\label{xsfig}
\end{figure}

Because the instantaneous CDF trigger rate might exceed our data handling 
capacity, the dimuon trigger, like many others, is prescaled. 
The integrated luminosity for the data sample has to be reduced by the 
luminosity-dependent dimuon trigger prescale factor to calculate the cross
section.
This is done on a run-by-run basis and has negligible statistical or 
systematic uncertainty.  The 1.1 fb$^{-1}$ sample luminosity is reduced to 
0.95 fb$^{-1}$ for this trigger.    
The resulting inclusive cross sections for prompt and $B$-decay production 
are listed in Table~\ref{xstab}. 
The prompt and $B$-decay data are plotted versus $p_T$ in Fig.~\ref{xsfig}(a). 
Data from the Run I CDF measurement~\cite{run1-xs} are also 
included in Fig.~\ref{xsfig}(b).

The major systematic uncertainties on these results are due to the systematic 
uncertainty in the luminosity determination (6\%)~\cite{lumi} and 
the polarization uncertainty in the acceptance 
calculation ($9\%$ at low $p_T$, $2\%$ at high $p_T$). 
Other systematic uncertainties arise from $p_T$ variations in 
the trigger ($< 3\%$) and reconstruction efficiencies ($4.7\%$).  
Systematic uncertainties due to the mass shape parametrization, 
fitting function parametrization, and prompt fraction determination are 
all less than $1\%$. 
The data in Fig.~\ref{xsfig} have both statistical and systematic uncertainties included.  

The integrated cross sections are calculated by summing the differential cross sections 
across $p_T$ bins. The systematic uncertainty on the integrated cross section is calculated 
by assuming that all sources of uncertainty are fully correlated among $p_T$ bins, 
with the exception of the trigger efficiency uncertainty, which is uncorrelated among $p_T$ 
bins. After calculating the uncertainty on the integrated cross section from each source, 
the total uncertainty is calculated by summing the individual contributions in quadrature. 

The integrated inclusive differential cross section for $p_T >$ 2 GeV/$c$ and 
$\mid y \mid < 0.6$ is measured to be:

\setlength{\extrarowheight}{0.2cm}
\begin{center}
\begin{tabular}{cc}
 $p_T(\psi(\mathrm{2S})) > 2 ~{\rm GeV}/c$\hspace*{1cm} & $\sigma(p\overline{p} \rightarrow \psi(\mathrm{2S}) X) \cdot \mathrm{Br} (\psi(\mathrm{2S})\rightarrow\mu^+ \mu^-)$ \\
 & $ = 3.29\pm 0.04 {\mathit (stat.)} \pm 0.32 {\mathit (syst.)} \ {\rm nb}$.\\
\end{tabular}
\end{center}

For comparison to the Run I measurement, 
we limit the $p_T$ range to $p_T >$ 5 GeV/$c$.
Then the measured integrated inclusive cross section is:

\begin{center}
\begin{tabular}{cc}
 $p_T(\psi(\mathrm{2S})) > 5 ~{\rm GeV}/c$ \hspace*{1cm} & $\sigma(p\overline{p} \rightarrow\psi(\mathrm{2S}) X )\cdot \mathrm{Br} (\psi(\mathrm{2S})\rightarrow\mu^+ \mu^-)$ \\
& $ = 0.69\pm 0.01 {\mathit (stat.)} \pm 0.06 {\mathit (syst.)} \ {\rm nb}$.\\
\end{tabular}
\end{center}

At 1.8 TeV the integrated inclusive cross section for $p_T >$ 5 GeV/$c$ 
and pseudorapidity $< 0.6$
was $0.57\pm0.04^{+0.08}_{-0.09}~{\rm nb}$~\cite{run1-xs}.  
The increase is (21 $\pm$ 19) \%, compared to a theoretical prediction of 
(14 $\pm$ 8) \% based on changes in the parton energy distribution at the 
higher collision energy~\cite{yellowbook}.  The uncertainty on the experimental ratio is dominated by the Run I measurement due to its much lower statistics.

\begin{figure}[th]
\begin{center}
\includegraphics[width=0.4\textwidth]{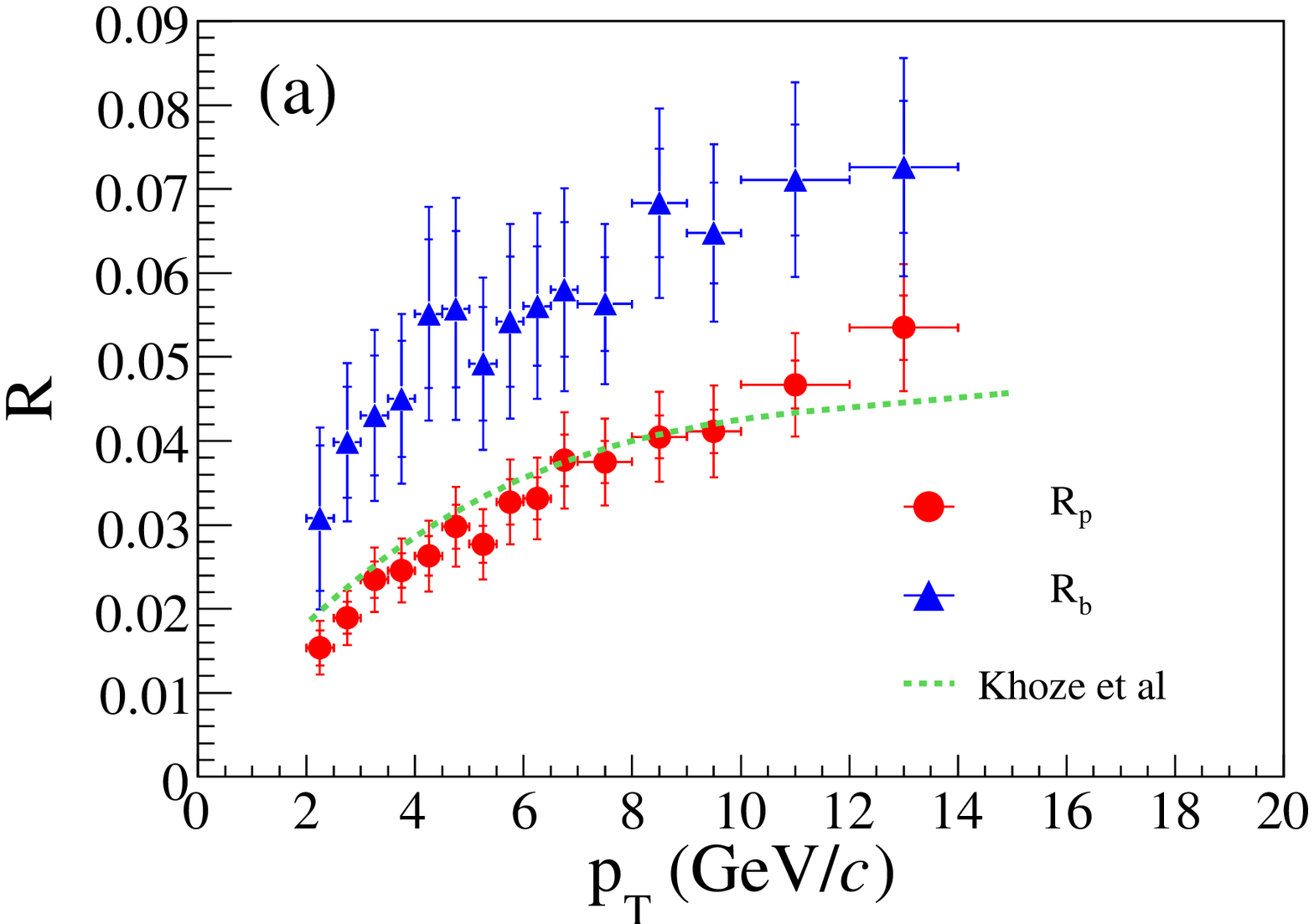}
\includegraphics[width=0.4\textwidth]{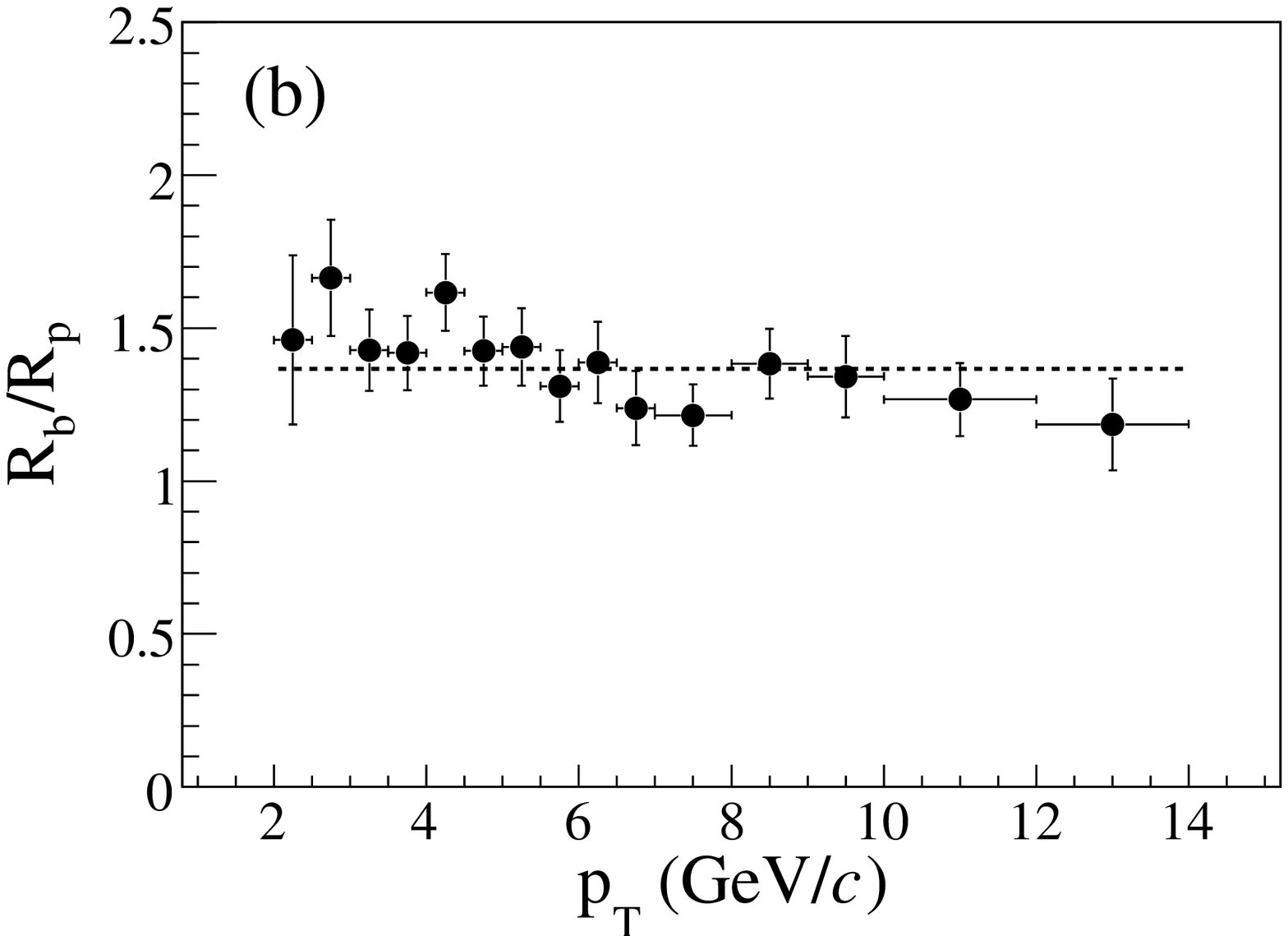}
\caption{(a) The differential cross section ratio of $\psi$(2S) to J/$\psi$ as
         a function of vector meson $p_T$ for prompt events ($R_p$) and 
         $B$-decay events ($R_b$). 	
	 The main error bar on each point is the quadrature sum of the statistical and 
	 uncorrelated systematic uncertainties for that $p_T$ bin.	
	The extensions to the error bars show the correlated uncertainty from the luminosity.  
	The dashed line is calculated using the prompt ratio predicted 
	from Ref.~\cite{gluon-tower} normalized to data 
	for the $p_T$ bin covering 8-9 GeV/$c$.  
         (b) The ratio of $B$-decay to prompt ratios $R_b/R_p$ as a function 
         of vector meson $p_T$.  The fit to a constant gives a $\chi^2$ of 
13 for 14 degrees of freedom.}
\label{comp}
\end{center}
\end{figure}

Prompt $\psi$(2S) production has a harder $p_T$ spectrum than that for
J/$\psi$ production~\cite{run2-xs}.  
We plot the ratio of these two cross sections as a function of vector meson
$p_T$ in Fig.~\ref{comp}(a). 
The increase in the ratio at larger $p_T$ reflects 
the slope difference.  Even though it neglects any feed-down contributions 
to J/$\psi$ prompt production, the model of Ref.~\cite{gluon-tower} predicts 
the $p_T$ dependence of this behavior, as shown by the dashed line in 
Fig.~\ref{comp}(a).  The model prediction is normalized to these data in the
$p_T$ bin covering 8-9 GeV/$c$.  
This same behavior is seen in the ratio of cross sections 
for production from $B$-decay, also shown in Fig.~\ref{comp}(a).  
The ratio of these two ratios is independent of $p_T$, 
as shown in Fig.~\ref{comp}(b).  
There is no theoretical motivation for this relation.

In conclusion, we have measured the $p_T$ dependence of the cross section for
$\psi$(2S) production in $p\overline{p}$ production at 1.96 TeV.  
These data have at least an order of magnitude more events than 
the Run I measurements and show more precisely the trends seen in those data. 
The increase in the inclusive cross section at the higher energy of Run II 
(1.96 TeV) compared to Run I (1.8 TeV) agrees
with expectations based on the increase in parton energy distribution.

These data extend the $\psi$(2S) differential cross section measurement up
to 30 GeV/$c$.  They are an important input for an update of the matrix 
elements in the NRQCD factorization approach~\cite{nrqcd}.
In the gluon tower model~\cite{gluon-tower}, the prompt hadroproduction of 
J/$\psi$, $\psi$(2S), and $\Upsilon$ states have been calculated.
The uncertainties of their calculation are rather large but 
their cross section prediction with adjusted parameters 
describes the published Tevatron data. 
In addition, their mechanism predicts a longitudinal polarization of J/$\psi$ 
at large transverse momentum which agrees qualitatively with the recent 
Tevatron measurement~\cite{run2-pol}. 
We hope that in future calculations with this and other models the 
uncertainties can be reduced in order to make a meaningful comparison to these
new cross section data.
A successful description of both the cross section data and polarization measurements 
in the perturbative $p_T$ region would demonstrate a good understanding of 
the charmonium hadroproduction mechanisms.

We thank the Fermilab staff and the technical staffs of the participating institutions for their vital contributions. This work was supported by the U.S. Department of Energy and National Science Foundation; the Italian Istituto Nazionale di Fisica Nucleare; the Ministry of Education, Culture, Sports, Science and Technology of Japan; the Natural Sciences and Engineering Research Council of Canada; the National Science Council of the Republic of China; the Swiss National Science Foundation; the A.P. Sloan Foundation; the Bundesministerium f\"ur Bildung und Forschung, Germany; the Korean Science and Engineering Foundation and the Korean Research Foundation; the Science and Technology Facilities Council and the Royal Society, UK; the Institut National de Physique Nucleaire et Physique des Particules/CNRS; the Russian Foundation for Basic Research; the Ministerio de Ciencia e Innovaci\'{o}n, and Programa Consolider-Ingenio 2010, Spain; the Slovak R\&D Agency; and the Academy of Finland. 


\end{document}

%% file: July_2008_Authors_Visitors.tex
\affiliation{Institute of Physics, Academia Sinica, Taipei, Taiwan 11529, Republic of China} 
\affiliation{Argonne National Laboratory, Argonne, Illinois 60439} 
\affiliation{University of Athens, 157 71 Athens, Greece} 
\affiliation{Institut de Fisica d'Altes Energies, Universitat Autonoma de Barcelona, E-08193, Bellaterra (Barcelona), Spain} 
\affiliation{Baylor University, Waco, Texas  76798} 
\affiliation{Istituto Nazionale di Fisica Nucleare Bologna, $^y$University of Bologna, I-40127 Bologna, Italy} 
\affiliation{Brandeis University, Waltham, Massachusetts 02254} 
\affiliation{University of California, Davis, Davis, California  95616} 
\affiliation{University of California, Los Angeles, Los Angeles, California  90024} 
\affiliation{University of California, San Diego, La Jolla, California  92093} 
\affiliation{University of California, Santa Barbara, Santa Barbara, California 93106} 
\affiliation{Instituto de Fisica de Cantabria, CSIC-University of Cantabria, 39005 Santander, Spain} 
\affiliation{Carnegie Mellon University, Pittsburgh, PA  15213} 
\affiliation{Enrico Fermi Institute, University of Chicago, Chicago, Illinois 60637}
\affiliation{Comenius University, 842 48 Bratislava, Slovakia; Institute of Experimental Physics, 040 01 Kosice, Slovakia} 
\affiliation{Joint Institute for Nuclear Research, RU-141980 Dubna, Russia} 
\affiliation{Duke University, Durham, North Carolina  27708} 
\affiliation{Fermi National Accelerator Laboratory, Batavia, Illinois 60510} 
\affiliation{University of Florida, Gainesville, Florida  32611} 
\affiliation{Laboratori Nazionali di Frascati, Istituto Nazionale di Fisica Nucleare, I-00044 Frascati, Italy} 
\affiliation{University of Geneva, CH-1211 Geneva 4, Switzerland} 
\affiliation{Glasgow University, Glasgow G12 8QQ, United Kingdom} 
\affiliation{Harvard University, Cambridge, Massachusetts 02138} 
\affiliation{Division of High Energy Physics, Department of Physics, University of Helsinki and Helsinki Institute of Physics, FIN-00014, Helsinki, Finland} 
\affiliation{University of Illinois, Urbana, Illinois 61801} 
\affiliation{The Johns Hopkins University, Baltimore, Maryland 21218} 
\affiliation{Institut f\"{u}r Experimentelle Kernphysik, Universit\"{a}t Karlsruhe, 76128 Karlsruhe, Germany} 
\affiliation{Center for High Energy Physics: Kyungpook National University, Daegu 702-701, Korea; Seoul National University, Seoul 151-742, Korea; Sungkyunkwan University, Suwon 440-746, Korea; Korea Institute of Science and Technology Information, Daejeon, 305-806, Korea; Chonnam National University, Gwangju,500-757, Korea; Chonbuk National University, Jeonju 561-756, Korea} 
\affiliation{Ernest Orlando Lawrence Berkeley National Laboratory, Berkeley, California 94720} 
\affiliation{University of Liverpool, Liverpool L69 7ZE, United Kingdom} 
\affiliation{University College London, London WC1E 6BT, United Kingdom} 
\affiliation{Centro de Investigaciones Energeticas Medioambientales y Tecnologicas, E-28040 Madrid, Spain} 
\affiliation{Massachusetts Institute of Technology, Cambridge, Massachusetts  02139} 
\affiliation{Institute of Particle Physics: McGill University, Montr\'{e}al, Qu\'{e}bec, Canada H3A~2T8; Simon Fraser University, Burnaby, British Columbia, Canada V5A~1S6; University of Toronto, Toronto, Ontario, Canada M5S~1A7; and TRIUMF, Vancouver, British Columbia, Canada V6T~2A3} 
\affiliation{University of Michigan, Ann Arbor, Michigan 48109} 
\affiliation{Michigan State University, East Lansing, Michigan  48824}
\affiliation{Institution for Theoretical and Experimental Physics, ITEP, Moscow 117259, Russia} 
\affiliation{University of New Mexico, Albuquerque, New Mexico 87131} 
\affiliation{Northwestern University, Evanston, Illinois  60208} 
\affiliation{The Ohio State University, Columbus, Ohio  43210} 
\affiliation{Okayama University, Okayama 700-8530, Japan} 
\affiliation{Osaka City University, Osaka 588, Japan} 
\affiliation{University of Oxford, Oxford OX1 3RH, United Kingdom} 
\affiliation{Istituto Nazionale di Fisica Nucleare, Sezione di Padova-Trento, $^z$University of Padova, I-35131 Padova, Italy} 
\affiliation{LPNHE, Universite Pierre et Marie Curie/IN2P3-CNRS, UMR7585, Paris, F-75252 France} 
\affiliation{University of Pennsylvania, Philadelphia, Pennsylvania 19104}
\affiliation{Istituto Nazionale di Fisica Nucleare Pisa, $^{aa}$University of Pisa, $^{bb}$University of Siena and $^{cc}$Scuola Normale Superiore, I-56127 Pisa, Italy} 
\affiliation{University of Pittsburgh, Pittsburgh, Pennsylvania 15260} 
\affiliation{Purdue University, West Lafayette, Indiana 47907} 
\affiliation{University of Rochester, Rochester, New York 14627} 
\affiliation{The Rockefeller University, New York, New York 10021} 
\affiliation{Istituto Nazionale di Fisica Nucleare, Sezione di Roma 1, $^{dd}$Sapienza Universit\`{a} di Roma, I-00185 Roma, Italy} 

\affiliation{Rutgers University, Piscataway, New Jersey 08855} 
\affiliation{Texas A\&M University, College Station, Texas 77843} 
\affiliation{Istituto Nazionale di Fisica Nucleare Trieste/Udine, I-34100 Trieste, $^{ee}$University of Trieste/Udine, I-33100 Udine, Italy} 
\affiliation{University of Tsukuba, Tsukuba, Ibaraki 305, Japan} 
\affiliation{Tufts University, Medford, Massachusetts 02155} 
\affiliation{Waseda University, Tokyo 169, Japan} 
\affiliation{Wayne State University, Detroit, Michigan  48201} 
\affiliation{University of Wisconsin, Madison, Wisconsin 53706} 
\affiliation{Yale University, New Haven, Connecticut 06520} 
\author{T.~Aaltonen}
\affiliation{Division of High Energy Physics, Department of Physics, University of Helsinki and Helsinki Institute of Physics, FIN-00014, Helsinki, Finland}
\author{J.~Adelman}
\affiliation{Enrico Fermi Institute, University of Chicago, Chicago, Illinois 60637}
\author{T.~Akimoto}
\affiliation{University of Tsukuba, Tsukuba, Ibaraki 305, Japan}
\author{B.~\'{A}lvarez~Gonz\'{a}lez$^t$}
\affiliation{Instituto de Fisica de Cantabria, CSIC-University of Cantabria, 39005 Santander, Spain}
\author{S.~Amerio$^z$}
\affiliation{Istituto Nazionale di Fisica Nucleare, Sezione di Padova-Trento, $^z$University of Padova, I-35131 Padova, Italy} 

\author{D.~Amidei}
\affiliation{University of Michigan, Ann Arbor, Michigan 48109}
\author{A.~Anastassov}
\affiliation{Northwestern University, Evanston, Illinois  60208}
\author{A.~Annovi}
\affiliation{Laboratori Nazionali di Frascati, Istituto Nazionale di Fisica Nucleare, I-00044 Frascati, Italy}
\author{J.~Antos}
\affiliation{Comenius University, 842 48 Bratislava, Slovakia; Institute of Experimental Physics, 040 01 Kosice, Slovakia}
\author{G.~Apollinari}
\affiliation{Fermi National Accelerator Laboratory, Batavia, Illinois 60510}
% ...

\collaboration{CDF Collaboration\footnote{With visitors from $^a$University of Massachusetts Amherst, Amherst, Massachusetts 01003,
$^b$Universiteit Antwerpen, B-2610 Antwerp, Belgium, 
$^c$University of Bristol, Bristol BS8 1TL, United Kingdom,
$^d$Chinese Academy of Sciences, Beijing 100864, China, 
$^e$Istituto Nazionale di Fisica Nucleare, Sezione di Cagliari, 09042 Monserrato (Cagliari), Italy,
$^f$University of California Irvine, Irvine, CA  92697, 
$^g$University of California Santa Cruz, Santa Cruz, CA  95064, 
$^h$Cornell University, Ithaca, NY  14853, 
$^i$University of Cyprus, Nicosia CY-1678, Cyprus, 
$^j$University College Dublin, Dublin 4, Ireland,
$^k$University of Edinburgh, Edinburgh EH9 3JZ, United Kingdom, 
$^l$University of Fukui, Fukui City, Fukui Prefecture, Japan 910-0017
$^m$Kinki University, Higashi-Osaka City, Japan 577-8502
$^n$Universidad Iberoamericana, Mexico D.F., Mexico,
$^o$University of Iowa, Iowa City, IA  52242,
$^p$Queen Mary, University of London, London, E1 4NS, England,
$^q$University of Manchester, Manchester M13 9PL, England, 
$^r$Nagasaki Institute of Applied Science, Nagasaki, Japan, 
$^s$University of Notre Dame, Notre Dame, IN 46556,
$^t$University de Oviedo, E-33007 Oviedo, Spain, 
$^u$Texas Tech University, Lubbock, TX  79609, 
$^v$IFIC(CSIC-Universitat de Valencia), 46071 Valencia, Spain,
$^w$University of Virginia, Charlottesville, VA  22904,
$^x$Bergische Universit\"at Wuppertal, 42097 Wuppertal, Germany,
$^{ff}$On leave from J.~Stefan Institute, Ljubljana, Slovenia, 
}}
\noaffiliation

%% file: psip_rc.bbl
\begin{thebibliography}{99}
\bibitem{CDF}
  The CDF coordinate system has $\hat{z}$ along the proton direction,
  $\hat{x}$ horizontal pointing outward from the Tevatron ring, and
  $\hat{y}$ vertical. $\theta$ ($\phi$) is the polar (azimuthal) angle
  measured with respect to $\hat{z}$ and $\eta$ is the pseudorapidity
  defined as $-{\rm ln}\;({\rm tan}\;(\theta/2))$. The transverse
  momentum of a particle is denoted as $p_T=p\;{\rm sin}\;\theta$.
\end{thebibliography}
